\documentclass[prd,a4paper,nofootinbib]{revtex4}

\usepackage{amsmath,amssymb}
\usepackage{epsfig}

\newcommand{\sigmadiff}{\sigma_{\mbox{\footnotesize diff}}}
\newcommand{\sigmatot}{\sigma_{\mbox{\footnotesize tot}}}
\newcommand{\sigmael}{\sigma_{\mbox{\footnotesize el}}}
\newcommand{\physr}{\rangle_{\mbox{\footnotesize phys}}}
\newcommand{\physl}{{}_{\mbox{\footnotesize phys}}\langle}

\newcommand{\pom}{{\mathbb P}}
\newcommand{\LambdaQCD}{\Lambda_{\mbox{\footnotesize QCD}}}

\begin{document}
\title{Diffractive photon dissociation
  in the saturation regime\\
 from the Good and Walker picture
}
\author{St{\'e}phane Munier%
}\email{Stephane.Munier@cpht.polytechnique.fr} 
\affiliation{Centre de physique th{\'e}orique%
\footnote{UMR 7644, unit{\'e} mixte de recherche du CNRS.},
{\'E}cole polytechnique, 91128 Palaiseau cedex, France}
\author{Arif Shoshi}\email{shoshi@phys.columbia.edu}
\affiliation{Department of Physics,
Columbia University, New York, NY 10027, USA.}

\begin{abstract}
Combining the QCD dipole model with the Good and Walker picture, 
we formulate
diffractive dissociation of a photon of virtuality $Q^2$ off a hadronic
target, in the kinematical regime in which
$Q$ is close to the saturation scale and much smaller than
the invariant mass of the diffracted system.
We show how the obtained formula compares to the HERA data 
and discuss what can be learnt from such a phenomenology.
In particular, we argue that diffractive observables in these
kinematics provide useful pieces of information on the
saturation regime of QCD.
\end{abstract}

\maketitle

\section{Introduction}

Hard diffraction has triggered a wide interest since its
discovery 
in high energy virtual photon-proton scattering
at HERA \cite{Derrick:1993xh}. It has been a
major theoretical challenge to understand the observed 
large rate of diffractive events within QCD,
especially for large virtualities $Q^2$ of the photon (for a
review, see Ref.~\cite{Hebecker:1999ej}).
Among the attempts to describe hard diffraction, the dipole
picture of QCD \cite{Nikolaev:1991ja,Mueller:1994rr} 
including unitarization corrections 
\cite{Golec-Biernat:1998js}
has proved particularly successful and natural in embedding 
both inclusive and diffractive observables in the 
same conceptual framework
\cite{Nikolaev:1991ja,Nikolaev:1992et,Mueller:1994jq,Bialas:1998vt}.

There are two distinct processes contributing to diffractive final
states. A first one consists in the photon splitting in a $q\bar q$ pair,
which scatters elastically off the target
without any further radiation.
The final state has typically a low invariant mass $M_X$, 
of the order of the virtuality of the photon. 
A second contribution is due to
the $q\bar q$ pair
interacting through higher Fock state fluctuations, for example 
$q\bar q g$, which go instead 
to a large-mass final state $M_X\gg Q$. 
The latter process is called
diffractive dissociation of the photon. As there is a clear
kinematical separation between the two processes, it makes sense to
analyse these contributions separately.
The present paper deals with 
the second one, {\it i.e.} diffractive dissociation.

To select this particular 
process one needs a large hierarchy between the mass of
the diffractive final state and the virtuality of the photon. This
means in practice that one is forced to relatively low virtuality scales
 $Q^2\!\leq\! 1\ \mbox{GeV}^2$, because of limited total 
center-of-mass energy available in the experiments.
However, the process can still be computable in perturbative QCD
provided that the energy is high enough to push the intrinsic $k_\bot$
of the partons inside the proton to large values 
$k_\bot\!\sim\! Q_s\!\gg\! \LambdaQCD$.
This feature is a consequence of the saturation of parton densities 
\cite{Gribov:1981ac}
and is now well-understood within QCD
\cite{Gribov:1981ac,McLerran:1994ni,
Balitsky:1996ub}.
$Q_s$ is called the saturation scale,
and is an increasing function of the rapidity of the proton.

The dipole picture is well-suited because color dipoles,
which are two-body color neutral objets characterized by their longitudinal
momentum, transverse size and impact parameter, are eigenstates of the
QCD interaction at high energy. Both elastic scattering (and
thus, from the optical theorem, the total cross section) 
and diffractive dissociation have a
straightforward formulation in terms of dipoles. The latter
is most simply obtained from the Good and Walker mechanism
\cite{Good:1960ba,Peschanski:1998cs}.

Our interest in diffractive dissociation 
is triggered by the wealth of new data
that are being taken in the relevant 
kinematical range at HERA. On the other hand, 
while some theoretical calculations are already available in the
literature \cite{Kovchegov:2001ni,Kovner:2001vi},
to our knowledge
there has been no phenomenological analysis of this particular
kinematical domain within QCD saturation models. 
Ref.~\cite{Golec-Biernat:1998js}
provides a model also for diffractive dissociation, but the
approximations made there allowed only for high
virtualities of the photon $Q\!\gg\! Q_s$, therefore the obtained
formulae are unapplicable to the recent data at low $Q^2$.

In the following,
we wish to clearly distinguish what is theoretically well under
control and which are the model assumptions 
that have to be made to come to the
description of the HERA data.
In this spirit, 
we will start by providing a complete derivation for the idealized process of
diffractive dissociation of a photon on a large nucleus, for which
the saturation corrections can be implemented in an easier way.
This will be done in Sec.~\ref{sec:2}.
Some analytical results are given in Sec.~\ref{sec:anres}.
In Sec.~\ref{sec:3}, we turn to phenomenology and explain
how we can adapt the obtained results to photon-proton reactions 
at HERA.
We compare our predictions to recently analyzed data.
Our conclusions are drawn in Sec.~\ref{sec:4}.


\section{Diffractive dissociation of a virtual photon 
on a large nucleus}
\label{sec:2}

In this section, we compute the cross section for diffractive
production of a hadronic system of large mass $M_X$ 
in a highly energetic
photon-nucleus collision. 
We limit ourselves to 
diffracted systems which have the following partonic content: $q\bar
q$ or $q\bar q g$. Higher Fock states are indeed not needed for large
but still moderate values of $M_X$.
The target is a large nucleus because, as will be discussed, complete
and numerically manageable results
can be obtained strictly speaking
only in this idealized framework.
We will also stick to the large-$N_c$ limit, and to the semi-classical
approximation implied by the high energy kinematics.

\begin{figure}
\begin{center}
\epsfig{file=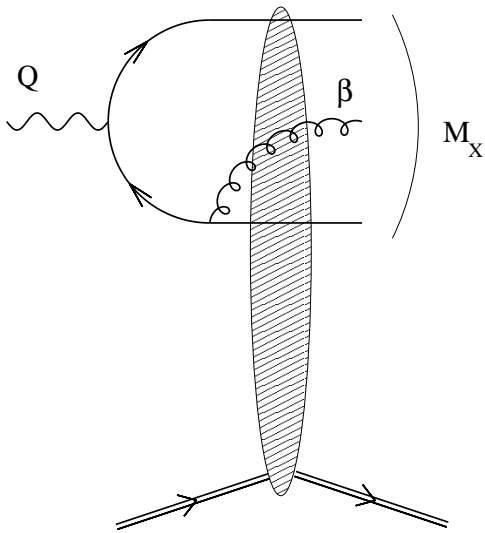,height=7cm}
\end{center}
\caption{One of the relevant graphs contributing to diffractive dissociation
  of the photon. The blob denotes the elastic double dipole-proton cross section.}
\end{figure}

\subsection{High energy kinematics and dipoles}

We go to a frame 
(the so-called dipole frame~\cite{Mueller:2001fv}) in which the 
photon and the nucleus have respective 4-momenta 
$q^\mu=(q^+,-Q^2/(2q^+),0^\bot)$ and
$p^{\mu}=(0,p^{-},0^\bot)$. The total
squared center-of-mass energy is given by $s=2q^+p^{-}\!-\!Q^2$.
Through an appropriate boost along the collision axis, 
the value of $q^{+}$ is chosen such that 
the whole diffracted partonic state $|q\bar q\rangle$ or 
$|q\bar qg\rangle$ is viewed as a quantum fluctuation of the photon
which subsequently scatters off the nucleus. The nucleus contains all
partonic fluctuations due to
the remaining energy evolution as part of its wave function at the
time of the interaction.
This is possible in a frame in which 
the rapidity of the slowest final-state particle is close to zero.
Of course, the observables will not depend on the frame, but
our choice provides a physical picture which is 
particularly adapted to
the study of unitarization effects.
The event is diffractive if the nucleus does not break up, which 
requires the scattering to proceed through the exchange of a
color-neutral object.

It was proven \cite{McLerran:1994ni,
Balitsky:1996ub} 
(for a review, see~\cite{Iancu:2003xm})
that in the dipole frame, the mean transverse momentum of the partons 
in the nucleus is shifted from $\LambdaQCD$ to a
scale $Q_s$ (the so-called saturation scale)
which increases with the rapidity of the nucleus,
due to multiple scatterings and saturation of the parton
densities \cite{Gribov:1981ac,Mueller:1990st} that
occur to preserve the unitarity of the scattering matrix. 
If the energy is large, then
the saturation scale can become high enough $Q_s\gg\LambdaQCD$
to justify perturbative
calculations for the diffracted partonic system, even if the 
transverse momenta at the level of the photon, 
initially of order $Q$, are not large.

In the dipole frame, the time sequence of the interaction
is the following: the photon
couples to a $q\bar q$ pair, which then radiates (or not) a gluon before
the whole system scatters off the nucleus through the
exchange of a color-neutral object. The leading terms at high
center-of-mass energy $s$ come about when the ordering in time is
strict, {\it i.e.} when the intermediate $q\bar q$ can be considered an almost
on-shell onium state. The following factorization then holds:
\begin{equation}
\frac{d\sigmadiff^{\gamma}}{d^2b}=\int d^2 r 
\int_0^1 dz
\left(|\psi_T^{\gamma}(r,z;Q)|^2
+|\psi_L^{\gamma}(r,z;Q)|^2
\right) \frac{d\sigmadiff}{d^2b}\ ,
\label{eq:facto}
\end{equation}
where $r$ is the transverse size of the $q\bar q$ pair, $b$ its impact
parameter, and $z$ is the longitudinal momentum fraction of the antiquark.
$d\sigmadiff/d^2b$ is the onium-nucleus diffractive cross section.
The photon wave functions $\psi^\gamma_T$, $\psi^\gamma_L$ on a $q\bar q$ pair 
are given by \cite{Bjorken:1971ah}
\begin{equation}
\begin{split}
|\psi^{\gamma}_T(r,z;Q)|^2&=
\frac{3\alpha_{em}}{2\pi^2}\sum_f e_f^2
\left((z^2+(1-z)^2)\varepsilon_f^2K_1^2(\varepsilon_f |r|)
+m_f^2 K_0^2(\varepsilon_f |r|)\right)\\
|\psi^{\gamma}_L(r,z;Q)|^2&=
\frac{3\alpha_{em}}{2\pi^2}\sum_f e_f^2
4Q^2 z^2(1-z)^2 K_0^2(\varepsilon_f|r|)
\end{split}
\label{eq:wf}
\end{equation}
for a transversely and longitudinally polarized photon respectively.
Here we sum over both polarizations, but in practice in the
kinematical domain of interest ($Q^2$ not too large), only the transverse
component will be relevant.
In the previous formula, $m_f$ is the mass of quark $f$ and 
\begin{equation}
\varepsilon_f=\sqrt{z(1-z)Q^2+m_f^2}\ .
\end{equation}
Thanks to the factorization~(\ref{eq:facto}), it will be enough to
compute the cross section $d\sigmadiff/d^2b$ for
diffractive dissociation of an {\it onium}. 
Using that formula, we will be able to relate it {\it in
  fine} in a straightforward way to the deep-inelastic scattering 
observables.

Recalling that $q^+$ is the lightcone momentum of the incident photon, 
the 4-momenta of the partons in its wave function read
\begin{equation}
\mbox{quark:}\ (z_q q^+,\frac{k_q^2}{2z_q q^+},k_q)\ ,\ 
\mbox{antiquark:}\ (z_{\bar q} q^+,\frac{k_{\bar q}^2}{2z_{\bar q}
  q^+},k_{\bar q})\ ,\
\mbox{gluon:}\ (z_{g} q^+,\frac{k_g^2}{2z_g
  q^+},k_g)
\end{equation}
where $z_q+z_{\bar q}+z_g=1$ and $k_q+k_{\bar q}+k_g=0$ from
conservation of the 3-momentum.
The transverse momenta $k_q$, $k_{\bar q}$ and $k_g$ 
are of the order of the external mass and/or virtuality scales
which we assume all of the same magnitude
$Q\gg\Lambda_{\mbox{\footnotesize QCD}}$.
We require furthermore the diffracted system to have a large mass
in the sense $M_X\gg Q$. Then the squared invariant mass of this
partonic system reads
\begin{equation}
M_X^2=Q^2\left(\frac{1}{z_q}+\frac{1}{z_{\bar q}}+\frac{1}{z_g}\right)
\end{equation}
and gets large if either $z_q$, $z_{\bar q}$ or $z_g$ are small. Due to the
infrared singularity of QCD, $z_g$ is very 
small in a typical configuration, 
in which case $M_X^2=Q^2/z_g$. This configuration with strong
ordering of longitudinal momenta $z_g\ll z_q,z_{\bar q}$ is the 
dominant contribution in the
kinematical regime of interest. Thus the mass of the diffractive final
state is directly related to the {\it longitudinal} 
momentum fraction of the photon
carried by the gluon.

Let us now introduce some more kinematics.
The difference between the rapidity of the onium and the rapidity of
the slowest particle in the final state is denoted by $y_X$. If
$\beta$ is the fraction of momentum of this particle with
respect to the initial-state onium, then 
it is easy to see that $y_X=\log(1/\beta)$. One also sees that $\beta=z_g$
for a $|q\bar qg\rangle$ final state.
The total rapidity difference between the photon and the nucleus is 
denoted by $y=\log(1/x_{Bj})$, where $x_{Bj}$ is the Bjorken variable.
Then one introduces the rapidity gap variable\footnote{%
Note that the saturation scale, which characterizes the transverse
momenta at the level of the nucleus, depends on $x_\pom$, and grows
when $x_\pom$ decreases.}
$y_\pom=y-y_X$ and similarly
to $\beta$, the variable
$x_\pom$ such that $y_\pom=\log(1/x_\pom)$.
The relationship between these variables and the kinematics of the
event reads
\begin{equation}
\beta=\frac{Q^2}{Q^2+M_X^2}\ ,\ \ \ x_{Bj}=\frac{Q^2}{Q^2+s}\ ,
\ \ \ x_\pom=\frac{x_{Bj}}{\beta}=\frac{Q^2+M_X^2}{Q^2+s}\ .
\end{equation}
The quantity which is measured and which will be computed is
the differential cross section
$d\sigmadiff^\gamma/dM_X$ or equivalently
$d\sigmadiff^\gamma/d\log(1/\beta)$. They are related
by a simple kinematical factor:
\begin{equation}
\frac{d\sigmadiff^\gamma}{dM_X}=
\frac{2M_X}{Q^2+M_X^2}
\frac{d\sigmadiff^\gamma}{d\log(1/\beta)}
\underset{M_X^2\gg Q^2}{=}
\frac{2}{M_X}\frac{d\sigmadiff^\gamma}{d\log(1/\beta)}\ .
\label{eq:obs}
\end{equation}

The calculation of the cross section for diffractive production of a
gluon in onium-nucleus scattering will proceed in two steps.
In the next subsection we compute the elastic scattering amplitude for
the onium on the nucleus up to order ${\cal O}(\alpha_s)$.
Then we will deduce the diffractive cross section 
from the Good and Walker formula and with the help of the
elastic amplitude.

\subsection{Elastic amplitude at next-to-leading order}

\begin{figure}
\begin{center}
\begin{equation}
\begin{array}{ccc}
\mbox{\epsfig{file=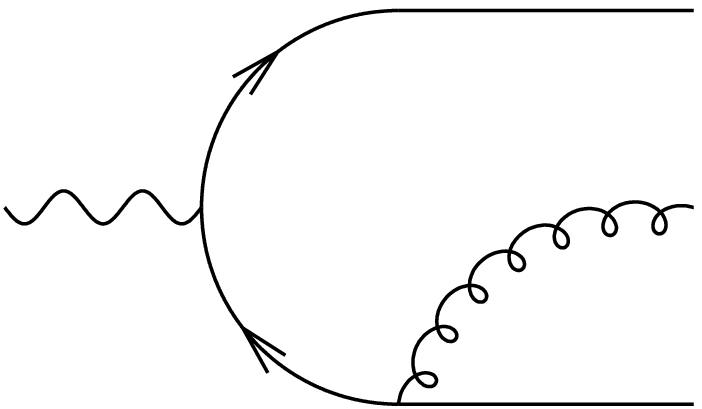,width=4cm}}&
\phantom{un grand espace}
&\mbox{\epsfig{file=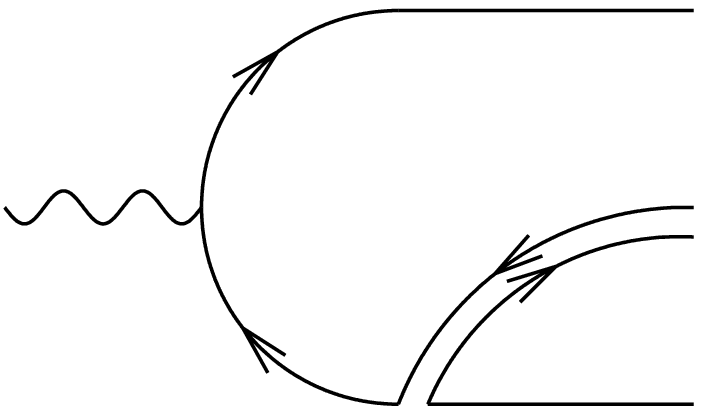,width=4cm}}\\
\\
\phantom{espace}\mbox{(a)}& &\phantom{espace}\mbox{   (b)}
\end{array}
\nonumber
\end{equation}
\end{center}
\caption{\label{fig:largeNc}(a) The $q\bar q g$ Fock state
  decomposition of the photon. 
(b) The same in the large-$N_c$ limit.}
\end{figure}

We only keep the leading terms in a $1/N_c$ expansion. Then, as
well-known, the partonic
content of a color neutral object can be represented as 
a set of dipoles (see Fig.~\ref{fig:largeNc}).

We denote by $x_i$ (a 2-dimensional vector)
the transverse position of the dipole constituent
$i$ with respect to the center of the target,
$x_{ij}=x_i-x_j$, and $z_i$ is the fraction of longitudinal momentum
carried by constituent $i$.
The normalization of a generic dipole state is given by the
orthogonality condition
\begin{equation}
\langle x,b,z| x^\prime,b^\prime,z^\prime\rangle=
\delta^2(x-x^\prime)\delta^2(b-b^\prime)\delta(z-z^\prime)
\label{eq:norm}
\end{equation}
for two dipoles of respective sizes $x$, $x^\prime$ and center-of-mass
coordinate vectors $b$, $b^\prime$.
In the following, unless it might hamper understanding,
one does not keep track of the $z$-variables in order to simplify the notations.

The physical dipole state $|x_{01},b\physr$ 
in the initial state of the experiment is dressed with all
possible fluctuations. Its expansion in terms of bare Fock states reads
\begin{equation}
|x_{01},b\physr
=\sqrt{Z(x_{01})}|x_{01},b\rangle+\int d^2x_2 dz_2\,
\psi(x_{02},x_{12},z_2)|x_{02},x_{12},b\rangle+\cdots
\label{eq:phys}
\end{equation}
where $\psi(x_{02},x_{12},z_2)$ is the probability amplitude that the
dipole $x_{01}$ split in two
dipoles of respective sizes $x_{02}$ and $x_{12}$, which corresponds to
a 1-gluon fluctuation of the initial onium. The gluon has longitudinal momentum fraction $z_2$.
Note that $|\psi|^2$ is of order $\alpha_s$, and the higher Fock states
not written are at least
of order $\alpha_s^2$: this remark provides the systematics for this
Fock-state expansion, seen as a perturbative expansion in the
parameter $\alpha_s$.
This expansion is relevant since $\alpha_s$ is small because it
runs with a scale between
$Q_s$ and $Q$, which we have both assumed much larger than $\LambdaQCD$.
Also, note that the 2-dipole state is not renormalized to this order,
but would get renormalized from higher orders.

The quantity that will appear in the calculation of observables is the
squared wave function, summed over the polarizations and colors of the
final-state gluon, whose expression can be found in \cite{Mueller:1994rr}:
\begin{equation}
\phi(x_{02},x_{12},z_2)\equiv\sum_{\mbox{\footnotesize
    polarization \& color}}|\psi(x_{02},x_{12},z_2)|^2
=\frac{\alpha_s N_c}{2\pi^2}\frac{1}{z_2}\frac{x_{01}^2}{x_{02}^2
    x_{12}^2}\ .
\end{equation}
From the unitarity condition~(\ref{eq:norm}) applied to the
physical state
$|x_{01},b\physr$ in~(\ref{eq:phys}), one can compute the
renormalization constant $Z$
\begin{equation}
Z(x_{01})=1-\int_{\rho^2}d^2 x_2\int_\beta^1 dz_2\,\phi(x_{02},x_{12},z_2)\ ,
\label{eq:renorm}
\end{equation}
where $\rho$ cuts off the ultraviolet divergences and $\beta$ the infrared
ones. In practice, as it should be $\rho$ disappears from physical quantities, and
$\beta$ is fixed by the kinematical limit.
Note that $Z$ represents physically the probability that the initial
onium does not radiate any gluon.

Let us denote by $\mathbb S$ the scattering matrix of the onium
on the nucleus.
Color dipoles form an orthogonal basis of the interaction eigenstates at
high energy, thus the ${\mathbb S}$-matrix element between two dipole
states reads
\begin{equation}
\langle x,b|{\mathbb S}|x^\prime,b^\prime\rangle
=\delta^2(x-x^\prime)\delta^2(b-b^\prime) S(x,b,x_\pom)\ .
\label{eq:smat}
\end{equation}
The quantity $|S(x,b,x_\pom)|^2$ is the probability that a dipole of
size $x$ at impact parameter $b$ does {\it not} interact with the
target.
The $x_\pom$-dependence of this quantity reflects the high energy
evolution of the target.

The ${\mathbb S}$-matrix element between two double dipole states reads
\begin{equation}
\langle x_{02},x_{12},b|{\mathbb S}|x^\prime_{02},x^\prime_{12},
b^\prime\rangle
=\delta^2(x_{02}-x^\prime_{02}) 
\delta^2(x_{12}-x^\prime_{12}) 
\delta^2(b-b^\prime) 
S(x_{02},x_{12},b,x_\pom)\ .
\label{eq:smat2dip}
\end{equation}
The notation is the following: $S(x_{02},x_{12},b,x_\pom)$ 
is the elastic diffusion matrix element
for states formed of two adjacent dipoles of respective
sizes $x_{02}$ and $x_{12}$, and $b=(x_0+x_1)/2$. The quantity
$|S(x_{02},x_{12},b,x_\pom)|^2$ is the probability that neither $x_{02}$ nor
$x_{12}$ interact, which under the assumption that the dipoles
interact independently 
is the product of the probabilities that $x_{02}$ and
$x_{12}$ do not interact, 
$|S(x_{02},b-x_{12}/2,x_\pom)|^2|S(x_{12},b-x_{02}/2,x_\pom)|^2$. 
The latter assumption is justified when the target is a large
nucleus, since nucleons are color-neutral and independent.
As furthermore ${\mathbb S}$ is known to be real at high energy, one
can get rid of the squared modulus and one gets
\begin{equation}
S(x_{02},x_{12},b,x_\pom)
=S(x_{02},b-x_{12}/2,x_\pom)S(x_{12},b-x_{02}/2,x_\pom)\ .
\label{eq:smatfact}
\end{equation}
Next one expresses the scattering matrix 
for the physical dipole $|x_{01},b\physr$ as
\begin{multline}
\physl x_{01},b|{\mathbb S}|x^\prime_{01},b^\prime\physr=
Z \langle x_{01},b|{\mathbb S}|x^\prime_{01},b^\prime\rangle\\
+\int_{\rho^2} d^2 x_2 d^2 x^\prime_2 \int_\beta^1 dz_2 dz_2^\prime\,
\psi^*(x_{02},x_{12},z_2)
\psi(x^\prime_{02},x^\prime_{12},z^\prime_2)\,
\langle x_{02},x_{12},b|{\mathbb
  S}|x^\prime_{02},x^\prime_{12},b^\prime\rangle\ .
\end{multline}
Using Eqs.~(\ref{eq:renorm},\ref{eq:smat},\ref{eq:smatfact}), 
and performing the integration over $x^\prime_{01}$ and $b^\prime$, one gets
for the diffusion amplitude
\begin{multline}
{\cal T}(x_{01},b,\beta,x_\pom)=T(x_{01},b,x_\pom)\\
+\frac{\alpha_s N_c}{2\pi^2}\int_\beta^1\frac{dz_2}{z_2}\int d^2 x_2\, 
\frac{x_{01}^2}{x_{02}^2 x_{12}^2} \left(S(x_{01},b,x_\pom)
-S(x_{02},b\!-\!x_{12}/2,x_\pom)S(x_{12},b\!-\!x_{02}/2,x_\pom)\right)\ ,
\label{eq:2dip}
\end{multline}
where $T$ stands for the matrix elements of ${\mathbb T}=1-{\mathbb
  S}$ between bare states, and ${\cal T}$ are its matrix elements
  betweeen dressed states.
The UV cutoff has been removed because when either $x_{02}$ or
$x_{12}$ tends to 0, then because of color transparency, the factor
built from the ${\mathbb S}$-matrix elements 
vanishes like $x^2_{02}$
(resp. $x^2_{12}$), and thus the integral
  over $x_2$ is finite.
The first term in Eq.~(\ref{eq:2dip}) is the Born term, 
the second one is the ${\cal O}(\alpha_s)$ correction to it. 
Physically, $\cal T$ is the amplitude for inclusive production of a
  forward gluon having longitudinal momentum larger than $\beta q^+$,
  when the total available rapidity is $y=\log(1/\beta x_\pom)$.

At this point, it is interesting to note that 
the procedure described above can be iterated to get
the Balitsky-Kovchegov (BK)
equation~\cite{Balitsky:1996ub}.
The latter gives the evolution of the dipole-nucleus ${\mathbb S}$-matrix
element for a dressed dipole $\hat S(x_{01},b,\beta,x_\pom)$ when the total
rapidity $y\!=\!\log(1/x_{Bj})\!=\!\log(1/\beta x_\pom)$ 
increases.
Starting from the scattering of a bare onium at $\beta\!=\!1$, Eq.~(\ref{eq:2dip})
gives the evolution of the amplitude ${\cal T}=1-\hat
S(x_{01},b,\beta,x_\pom)$ as one decreases $\beta$, {\it i.e.} as one
gives a
small boost to the onium while the rapidity of the target is kept fixed.
The change of the amplitude comes from the fact that the onium can now
be found with an extra gluon in its wave function: in the large $N_c$
limit, this means that the onium has a probability to be formed of
{\it two} adjacent dipoles.
Then under the
assumption that the dipoles interact {\it independently}, for example with
different nucleons of a large nucleus, this procedure can be iterated
by simply replacing the ${\mathbb
  S}$-matrix element for {\it bare} dipoles in the r.h.s. of
Eq.~(\ref{eq:2dip}) (for which $\beta\!=\!1$) by the ${\mathbb
  S}$-matrix element for the scattering of {\it dressed} dipoles at
total rapidity
$\log(1/z_2x_\pom)<\log(1/\beta x_\pom)$, which reads 
$\hat S(x_{01},b,z_2,x_\pom)$. 
One gets the following integral equation:
\begin{multline}
\hat S(x_{01},b,\beta,x_\pom)=S(x_{01},b,x_\pom)\\
+\frac{\alpha_s N_c}{2\pi^2}
\int_\beta^1\frac{dz_2}{z_2}
\int d^2 x_2\, \frac{x_{01}^2}{x_{02}^2 x_{12}^2}
 \left(\hat S(x_{02},b-x_{12}/2,z_2,x_\pom)
\hat S(x_{12},b-x_{02}/2,z_2,x_\pom)-\hat S(x_{01},b,z_2,x_\pom)\right)\ .
\end{multline}
One now takes the derivative of both sides with respect to
$\log(1/\beta)$.
As we keep $x_\pom$ fixed, this is equivalent to taking the derivative
with respect to the total rapidity $\log(1/x_{Bj})$ and one
gets\footnote{%
We suppress the dependence upon the two variables $\beta$, $x_\pom$
in Eq.~(\ref{bk}) because an inclusive observable is only
sensitive to $x_{Bj}=\beta x_\pom$. Our derivation holds
in a specific frame defined by the rapidity sharing between
the onium and the target given by $\beta$,
$x_\pom$, but of course the physical observable 
computed here is independent of this choice.}
\begin{equation}
\frac{\partial {\hat S}(x_{01},b,x_{Bj})}{\partial \log(1/x_{Bj})}
=\frac{\alpha_s N_c}{2\pi^2}\int d^2 x_2\, \frac{x_{01}^2}{x_{02}^2 x_{12}^2}
 \left(\hat S(x_{02},b-x_{12}/2,x_{Bj})
\hat S(x_{12},b-x_{02}/2,x_{Bj})-\hat S(x_{01},b,x_{Bj})\right)\ ,
\label{bk}
\end{equation}
which is indeed the BK equation~\cite{Balitsky:1996ub} (see
Ref.~\cite{Mueller:2001fv} for a derivation closely related to ours).
This equation resums the so-called ``fan diagrams''.

Coming back to Eq.~(\ref{eq:2dip}) for ${\cal T}$,
the total cross section stems from the optical theorem, 
and the elastic cross section is the
squared amplitude ${\cal T}$:
\begin{equation}
\frac{d\sigmatot}{d^2b}=2{\cal T}(x_{01},b)\ \ \ \mbox{and}\ \ \
\frac{d\sigmael}{d^2b}=|{\cal T}(x_{01},b)|^2\ .
\label{eq:formtotel}
\end{equation}

\subsection{Diffraction from the Good and Walker picture}

An elegant and most straightforward 
formulation of diffractive dissociation is obtained by using
the Good and Walker~\cite{Good:1960ba} picture (see
also~\cite{Miettinen:1978jb} and~\cite{Peschanski:1998cs}), according to which
inelastic diffraction is
proportional to the dispersion in cross sections for the diagonal
channels of the scattering. 
The reason why we are relying on such a picture, 
developed in the context of early
hadronic physics \cite{Good:1960ba} and subsequently
extended to the parton model \cite{Miettinen:1978jb}, is that the
dipoles form precisely 
a complete set of eigenstates of the QCD interaction at
high energy.
The cross section for diffractive dissociation reads
in this picture
\begin{equation}
\frac{d\sigmadiff}{d^2b}=\int d^2 x^\prime_{01} 
d^2 b^\prime
\sum_X
\physl x_{01},b|
{\mathbb T}^\dagger |X\rangle\langle X| {\mathbb T}
|x^\prime_{01},b^\prime\physr
-\left|\int d^2 x^\prime_{01} 
d^2 b^\prime
\physl x_{01},b|
{\mathbb T}
|x^\prime_{01},b^\prime\physr\right|^2\ ,
\label{eq:goodwalker}
\end{equation}
where $X$ is any dipole final state, and the sum over $X$ also
contains an integration over phase space of the type $\int\prod d^2x_a dz_a$.
One denotes by (I) and (II) the (positive) terms appearing in this equation.

The second term is the easiest to compute since it
is simply ${d\sigmael}/{d^2b}$
(see formula~(\ref{eq:formtotel})). Keeping only the terms relevant at
order ${\cal O}(\alpha_s)$, one gets
\begin{equation}
\mbox{(II)}
=\frac{d\sigmael}{d^2b}
=T^2(x_{01},b,x_\pom)+\frac{\alpha_s N_c}{\pi^2}
\int_\beta^1\frac{dz_2}{z_2}
\int d^2x_2\frac{x_{01}^2}{x_{02}^2 x_{12}^2}
(1-S(x_{01}))(S(x_{01})-S(x_{02})S(x_{12}))\ .
\end{equation}
Coming back to the first term in Eq.~(\ref{eq:goodwalker}), 
the sum over $X$ is restricted to single $|x_{ab};b\rangle$ and double
$|x_{ac},x_{bc};b\rangle$
dipole states, since we limit ourselves to the order ${\cal
  O}(\alpha_s)$ of perturbation theory.
At high energy 
the interaction with the target cannot change the number of dipoles in the wave
function, the first term thus gives
\begin{multline}
\mbox{(I)}=\int d^2x^\prime_{01} d^2b^\prime\bigg\{Z(x_{01})
\int d^2x_{ab} d^2b^{\prime\prime} \langle x_{01},b|
{\mathbb T}^\dagger |x_{ab},b^{\prime\prime}\rangle\langle x_{ab},b^{\prime\prime}| {\mathbb T}
|x^\prime_{01},b^\prime\rangle\\
+\int d^2x_2 d^2x^\prime_2 d^2x_{ac} d^2 x_{bc} d^2b^{\prime\prime}
\psi^\dagger(x_{02},x_{12})\psi(x^\prime_{02},x^\prime_{12})\\
\times\langle x_{02},x_{12},b|
{\mathbb T}^\dagger |x_{ac},x_{bc},b^{\prime\prime}\rangle
\langle x_{ac},x_{bc},b^{\prime\prime}
| {\mathbb T}
|x^\prime_{02},x^\prime_{12},b^\prime\rangle\bigg\}\ .
\end{multline}
Eq.~(\ref{eq:renorm}) enables one to express $Z(x_{01})$ and 
Eqs.~(\ref{eq:smat2dip},\ref{eq:smatfact}) to replace the matrix
elements. Then one uses the $\delta$-functions coming in
Eqs.~(\ref{eq:smat2dip},\ref{eq:smatfact}) which are due to the fact
that the dipoles are eigenstates of the interaction, 
to perform the integrations over $x^\prime_{01}$, 
$x^\prime_2$, $x_{ab}$, $x_{ac}$, $x_{bc}$, $b^\prime$,
$b^{\prime\prime}$. One finds
\begin{equation}
\mbox{(I)}=T^2(x_{01})+\frac{\alpha_s N_c}{2\pi^2}
\int_\beta^1\frac{dz_2}{z_2}
\int d^2x_2\frac{x_{01}^2}{x_{02}^2 x_{12}^2}
\left\{
(1-S(x_{02})S(x_{12}))^2-(1-S(x_{01}))^2
\right\}\ .
\end{equation}
Putting the two parts (I) and (II)
together and taking the derivative with respect
to $\log(1/\beta)$, 
one gets after some straightforward algebra
\begin{equation}
\frac{d\sigmadiff}{d^2b\, d\log(1/\beta)}=\frac{\alpha_s N_c}{2\pi^2}
\int d^2 x_2 \frac{x_{01}^2}{x_{02}^2
  x_{12}^2}(S(x_{02},b-x_{12}/2,x_\pom)
S(x_{12},b-x_{02}/2,x_\pom)-S(x_{01},b,x_\pom))^2\ .
\label{eq:sigmadiff}
\end{equation}
A few technical remarks are in order.
Our result is identical to the one obtained in earlier derivations within
different frameworks (semi-classical~\cite{Kovchegov:2001ni}, 
eikonal~\cite{Kovner:2001vi}).
It is also consistent, in the weak interaction limit in which
$T=1-S\ll 1$, with previous
calculations~\cite{Nikolaev:1994th}, and in particular with the
computation of all Feynman diagrams contributing to the $q\bar q g$
final state~\cite{Bartels:1999tn}.

Diagrams 
with gluons emitted in the final state are not included in our calculation
since the radiative corrections we have concern exclusively 
the wave-function: no final state
interaction can be accounted for in our calculation. This class of
diagrams contributes to the diffractive final state, but they
cancel out if one does not measure the transverse momentum of the
gluon (see \cite{Chen:1995pa,Mueller:2001fv}).
Note that we could not obtain more detailed properties of the final
state within our derivation, but as demonstrated earlier, thanks to the
selected kinematics we do not
need them for our purpose.

Using the factorization~(\ref{eq:facto}) and the kinematical
relation~(\ref{eq:obs}), one gets the cross section for diffractive
dissociation of a virtual photon in the form
\begin{multline}
\frac{d\sigmadiff^\gamma}{dM_X}=\frac{\alpha_s N_c}{\pi^2}
\frac{1}{M_X}
\int d^2 x_{01}\int_0^1 dz
\left(|\psi_T^\gamma(z,x_{01};Q)|^2+
|\psi_L^\gamma(z,x_{01};Q)|^2\right)\\
\times\int d^2x_2 \frac{x_{01}^2}{x_{02}^2 x_{12}^2}
\int d^2b\,(S(x_{02},b-x_{12}/2,x_\pom)
S(x_{12},b-x_{02}/2,x_\pom)-S(x_{01},b,x_\pom))^2\ .
\label{eq:sigmadiffgamma}
\end{multline}
Equation~(\ref{eq:sigmadiffgamma}) has to be evaluated numerically for a
general dipole-nucleus ${\mathbb S}$-matrix, but there are two interesting limits
that can be studied analytically, namely $Q^2\gg Q_s^2$ and $Q^2
\ll Q_s^2$.
This is done in the next section.


\section{Analytical insight}
\label{sec:anres}

 We will need a few general properties of the ${\mathbb S}$-matrix
element:
\begin{alignat}{2}
S(r)&\underset{|r|\gg 1/Q_s}{\ll}1&\quad&\text{(black disk limit)}\\ 
1-S(r)&\underset{|r|\ll 1/Q_s}{\sim} Q_s^2 r^2& &\text{(color
  transparency)}\ .
\label{eq:smatapprox}
\end{alignat}
Let us also recall that this theoretical study assumes
enough center-of-mass energy so that $Q_s^2\gg\LambdaQCD^2$
in the dipole frame. This also means
that the dipole sizes entering formula~(\ref{eq:sigmadiff}), 
of order $1/Q_s$, are always smaller than the typical scale
$1/\LambdaQCD$ of the
variations of the ${\mathbb S}$-matrix as a function of the
impact parameter. Thus $S(x_{02},b-x_{12}/2)\simeq S(x_{02},b)$ and 
$S(x_{12},b-x_{02}/2)\simeq S(x_{12},b)$. We will not 
have to take care anymore of the $b$-dependence in the ${\mathbb
  S}$-matrix elements, because it factorizes completely.

\subsection{Onium-nucleus cross section}

\subsubsection{Small onium: collinear region}

We now analyse formula~(\ref{eq:sigmadiff}) in the limit $x_{01}^2 Q_s^2\ll 1$,
$S(x_{01})\simeq 1$. The dipole splitting probability is singular
for either $|x_{02}|\ll |x_{01}|$ or $|x_{12}|\ll |x_{01}|$. Let us consider
the first case, the second one being exactly symmetric. 
The ordering relation implies $|x_{12}|\sim |x_{01}|$. The contribution of
these dipole configurations to the cross section 
in Eq.~(\ref{eq:sigmadiff})
can then be rewritten as
\begin{equation}
\frac{\alpha_s N_c}{2\pi^2}
\int^{x_{01}^2} d^2 x_2 \frac{1}{x_{02}^2}(S(x_{02})-1)^2\ .
\end{equation}
In this domain one has $1-S(x_{02})\sim Q_s^2 x_{02}^2$,
so the integral vanishes as $Q_s^4 x_{01}^4$.

\begin{figure}
\begin{center}
\begin{equation}
\begin{array}{ccc}
\mbox{\epsfig{file=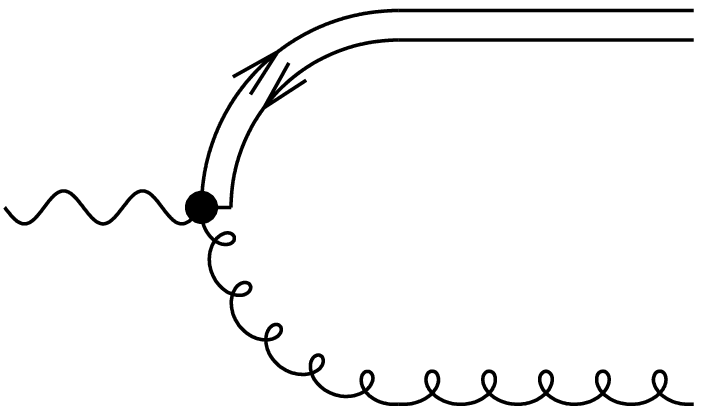,width=4cm}}&
\phantom{un grand espace}
&\mbox{\epsfig{file=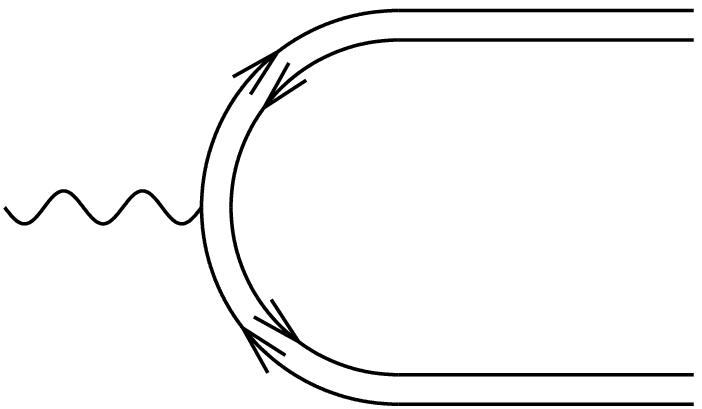,width=4cm}}\\
\\
\phantom{espace}\mbox{(a)}& &\phantom{espace}\mbox{   (b)}
\end{array}
\nonumber
\end{equation}
\end{center}
\caption{\label{fig:largeQ2}(a) The diffractive final state for $Q\gg Q_s$. 
(b) The same in the large-$N_c$ limit.}
\end{figure}

A second kinematical region which could bring large contributions to
the integral over $x_2$ is $|x_{02}|\gg |x_{01}|$,
which implies $|x_{02}|\sim |x_{12}|$. Then the contribution of these
configurations to formula~(\ref{eq:sigmadiff}) is
\begin{equation}
\frac{\alpha_s N_c}{2\pi^2}
\int_{x_{01}^2} d^2 x_2 \frac{x_{01}^2}{x_{02}^4}(S^2(x_{02})-1)^2\ .
\label{eq:contrib2}
\end{equation}
Using Eq.~(\ref{eq:smatapprox}), one sees that this integral
is of order $Q_s^2 x_{01}^2$ and thus this kinematical region
dominates over the previous one. 
It corresponds to configurations for which the $q\bar q$ pair is small and
well separated from the gluon, or, in momentum space, the gluon has a
small transverse momentum with respect to the one of the quark 
(see Fig.~\ref{fig:largeQ2}). 
This is the collinear limit.
The lower boundary of the integral~(\ref{eq:contrib2}) 
can be put to zero since the
region $|x_{02}|\!<\!|x_{01}|$ gives a subleading contribution.
The result reads most generally
\begin{equation}
\frac{d\sigmadiff}{d^2b\, d\log(1/\beta)}=
\frac{\alpha_s N_c}{2\pi^2}x_{01}^2
\int \frac{d^2 x}{x^4}(1-S^2(x))^2\ .
\label{eq:approxlargeq}
\end{equation}
The integral appearing here is proportional to $Q_s^2$. The
value of the associated proportionality constant
has to be computed in specific models for $S$.

The interpretation of the factor $(1-S^2(x,b))^2$ is the following.
$S^2(x,b)$ is the probability that a $q\bar q$ dipole of size $x$ at
impact parameter $b$ does not interact with the nucleus. The
elastic cross section reads in this case $d\sigmael/d^2b=(1-S(x,b))^2$. In the
same way, $S^4(x,b)$
would be the probability that {\it two} independent but superimposed
dipoles do not interact, and $d\sigmael/d^2b=(1-S^2(x,b))^2$ would be
the corresponding cross section. So the quantity $(1-S^2(x,b))^2$ can be understood as
the elastic cross section for an {\it octet} dipole, because in the large $N_c$ limit
the latter is represented by two independent lines of indices.

\subsubsection{Large onium: saturation region}

\begin{figure}[th]
\begin{center}
\epsfig{file=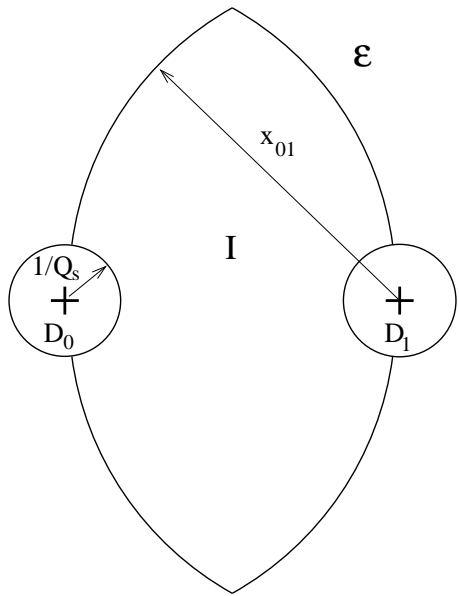,width=6cm}
\end{center}
\caption{\label{zone}The transverse plane and the different
  integration domains ${\cal D}_0$, ${\cal D}_1$,
${\cal I}$ and ${\cal E}$. The crosses denote the positions of the quark
  and the antiquark, $x_0$ and $x_1$ resp.}
\end{figure}

We now turn to the limit $x_{01}^2 Q_s^2\gg 1$. The discussion is
more subtle in this case because unlike the previous case
one cannot identify a typical
dipole configuration which gives the dominant contribution to the
cross section.
Let us split the integration domain into several relevant subdomains
as shown in Fig.~\ref{zone}. 
First consider the
disks ${\cal D}_0$ and
${\cal D}_1$ of radius $1/Q_s$ around
$x_0$ and $x_1$ respectively, inside of which 
$S(x_{12})\sim S(x_{01})$ and $S(x_{02})\sim
1$ (resp. $S(x_{02})\sim S(x_{01})$ and $S(x_{12})\sim 1$). Next define the
domain $\cal I$ for which either $|x_{02}|$ or $|x_{12}|$ is smaller than 
$|x_{01}|$ (the intersection with ${\cal D}_0$ and ${\cal D}_1$ is
excluded). The remaining part of the 2-dimensional plane is denoted by
$\cal E$. 

In the latter integration region $\cal E$, necessarily $S(x_{02})S(x_{12})\ll
S(x_{01})$ but in $\cal I$, the ordering relation depends on the model
for $S$.
Taking the relevant approximations in these different
domains, the integral in Eq.~(\ref{eq:sigmadiff}) then reads
\begin{equation}
S^2(x_{01})\int_{{\cal D}_0+{\cal D}_1}d^2x_2 \left(\frac{\partial
    S}{\partial x_{02}}(x_{02})\right)^2
+S^2(x_{01})\int_{{\cal E}}{d^2x_2}\frac{x_{01}^2}{x_{02}^2 x_{12}^2}
+\int_{{\cal I}}d^2x_2\frac{x_{01}^2}{x_{02}^2 x_{12}^2}
(S(x_{02})S(x_{12})-S(x_{01}))^2\ .
\end{equation}
The first term goes like $S^2(x_{01})$, and the second term is enhanced by a
factor of $\log(Q_s^2 x_{01}^2)$. 
As for the third term, no completely general statement
can be made. If $S(x_{02})S(x_{12})\ll
S(x_{01})$ everywhere in domain $\cal I$, the third term contributes
as much as the second term. 
Then the cross section goes like $S^2(x_{01})\log(x_{01}^2 Q_s^2)$.
This is the case for color glass condensate for which 
$S(x_{01})\sim \exp(-\lambda\log^2(x_{01}^2 Q_s^2))$, with
$\lambda\sim 0.2$ \cite{Levin:1999mw,Iancu:2002aq}. 
If $S(x_{02})S(x_{12})\geq S(x_{01})$ instead in some region
inside $\cal I$ (and in particular at the point 
$|x_{02}|\!=\!|x_{12}|\!=\!|x_{01}|/2$), 
the third term contributes as
$S^4(x_{01}/2)/x_{01}^2$, and can even dominate over the two other
ones. This is what happens when $S$ is given by the Golec-Biernat and
W{\"u}sthoff model.
Thus we see that the diffractive cross section behaves
{\it qualitatively} differently for the various models for the dipole-nucleus
${\mathbb S}$-matrix. To summarize, in these two kinds of model one
finds the following estimates for the asymptotic behaviors:
\begin{equation}
\frac{d\sigmadiff}{d^2b\, d\log(1/\beta)}
\simeq\begin{cases}
{\displaystyle \frac{\alpha_s N_c}{\pi}}
S^2(x_{01})\log(x_{01}^2 Q_s^2) &\text{for CGC,} \\
{\displaystyle \frac{4\alpha_s N_c}{\pi}}{S^4(x_{01}/2)}/{x_{01}^2}  
  &\text{for GBW, up to a ${\cal O}(1)$ factor.}
\end{cases}
\label{eq:approxsmallq}
\end{equation}
For GBW-like models, ${\cal O}(1)$ factors are cumbersome to get.

The onium cross section 
is anyway strongly damped in this region of
 large $x_{01}^2Q_s^2$ because $S(x_{01})$ goes to 0
in this limit, much faster 
than $1/(\log(x_{01}^2 Q_s^2))$ in all available
models. This is the black disc limit, in which the
elastic cross section is half the total one, and thus the
diffractive cross section is zero from the inequality
\begin{equation}
\frac{d\sigmadiff}{d^2b}\leq
\frac12\frac{d\sigmatot}{d^2b}-\frac{d\sigmael}{d^2b}\ ,
\end{equation}
which directly follows from unitarity \cite{Miettinen:1978jb}. 

To summarize, we see that the diffractive cross section increases with
$|x_{01}|$ like
$x_{01}^2$ at small $x_{01}^2 Q_s^2$ and decreases strongly
at large $x_{01}^2 Q_s^2$, so there must be a maximum 
for $x_{01}^2\sim 1/Q_s^2$.
For a realistic target, the saturation scale is a decreasing function
of the impact parameter. For a well-chosen value of $x_{01}$, 
it can happen that $Q_s^2(b)
x_{01}^2\gg 1$ for small $b$, 
and $Q_s^2(b) x_{01}^2\ll 1$ for large $b$ (this is always the
case). In such a situation, the main contribution to the diffractive
cross section for an incident dipole of size $|x_{01}|$ comes from a
zone in impact parameter which has the shape of a ring of radius $b_0$
such that  $Q_s^2(b_0)\sim 1/x_{01}^2$.
We have recovered the well-known fact \cite{Miettinen:1978jb}
that diffractive dissociation is peripheral at variance with elastic
or total scattering. However, the ``periphery'' of the nucleus is
defined with the help of the saturation scale and not with $\LambdaQCD$
at variance with soft processes.

\begin{figure}
\epsfig{file=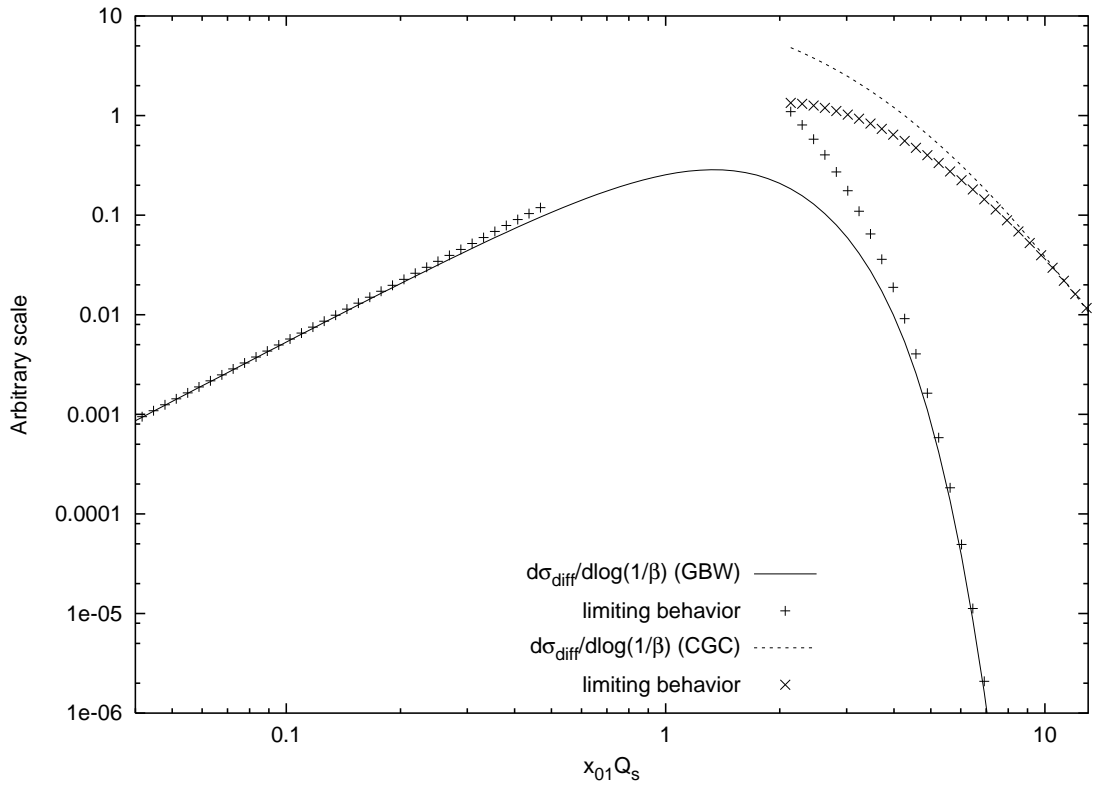,width=15cm}
\caption{\label{fig:plotapprox}The diffractive 
cross section compared to the 
approximations in extreme kinematical regimes.}
\end{figure}

The features just found are illustrated on Fig.~\ref{fig:plotapprox}
for the Golec-Biernat and W{\"u}sthoff model, and for the color glass
condensate in the large $x_{01}^2Q_s^2$ limit.

\subsection{Limiting behaviors of the photon-nucleus cross section}
\label{sec:lim}

Before we turn to the comparison to the data, we wish to study
some analytical properties of the cross
section~(\ref{eq:sigmadiffgamma}) 
in the two different limits
$Q^2\gg Q_s^2$ and $Q^2\ll Q_s^2$.
We focus here on transversely polarized photons.
To study asymptotic behaviors, it is enough to replace the photon wave
function~(\ref{eq:wf}) that appears in Eq.~(\ref{eq:sigmadiffgamma}) 
by the following approximation:
\begin{equation}
|\psi^{\gamma}_T(x_{01},z;Q)|^2\simeq
\frac{3\alpha_{em}}{2\pi^2}\sum_f e_f^2
(z^2+(1-z)^2)\frac{1}{x_{01}^2}\Theta(1-z(1-z)Q^2 x_{01}^2)\ .
\label{eq:aljets}
\end{equation}
Let us first treat the integration over $z$ in
Eq.~(\ref{eq:sigmadiffgamma}). There are two cases to
distinguish. If $Q^2x_{01}^2\!<\!4$ then the
$\Theta$-function in~(\ref{eq:aljets}) can be ignored and
\begin{equation}
\int_0^1 dz\,|\psi^{\gamma}_T(x_{01},z;Q)|^2=
\frac{\alpha_{em}}{\pi^2}\sum_f e_f^2
\frac{1}{x_{01}^2}\ .
\label{eq:approxwf1}
\end{equation}
If $Q^2x_{01}^2\!>\!4$ instead, then the leading contribution, obtained for 
$Q^2x_{01}^2\!\gg\! 4$, comes from the endpoints
\begin{equation}
\begin{split}
\int_0^1 dz\,|\psi^{\gamma}_T(x_{01},z;Q)|^2&=
\int_0^{1/(Q^2 x_{01}^2)} dz \,|\psi^{\gamma}_T(x_{01},z;Q)|^2
+\int_{1-1/(Q^2 x_{01}^2)}^1 dz \,|\psi^{\gamma}_T(x_{01},z;Q)|^2\\
&=\frac{3\alpha_{em}}{\pi^2}\sum_f e_f^2\frac{1}{Q^2 x_{01}^4}\ .
\label{eq:approxwf2}
\end{split}
\end{equation}
This is the well-known aligned-jet configuration \cite{Bjorken:1972ru}.


As for the case $Q^2\gg Q_s^2$, 
the integral over $x_{01}$ can be split in 3 integration regions
$[0,4/Q^2]$, $[4/Q^2,1/Q_s^2]$ and $[1/Q_s^2,\infty]$.
Taking the relevant approximations for the wave 
function~(\ref{eq:approxwf1}), (\ref{eq:approxwf2}) and for the
dipole ${\mathbb S}$-matrix
element~(\ref{eq:approxlargeq}), (\ref{eq:approxsmallq}) 
in each of these domains, one sees easily that region
$[4/Q^2,1/Q_s^2]$ dominates. 
Eq.~(\ref{eq:sigmadiffgamma}) reduces to
\begin{equation}
\frac{d\sigmadiff^{\gamma}}{dM_X}=
\frac{6\alpha_{em}N_c}{\pi^2 M_X}\frac{1}{Q^2}
\sum e_f^2\frac{\alpha_s}{2\pi}
\int d^2b\,
\log\frac{Q^2}{Q_s^2}\int\frac{d^2x}{x^4}(1-S^2(x,b,x_\pom))^2\ .
\label{eq:approxlargeqcs}
\end{equation}
This limit is consistent with previously derived results, 
as shown in the appendix.

\begin{figure}
\epsfig{file=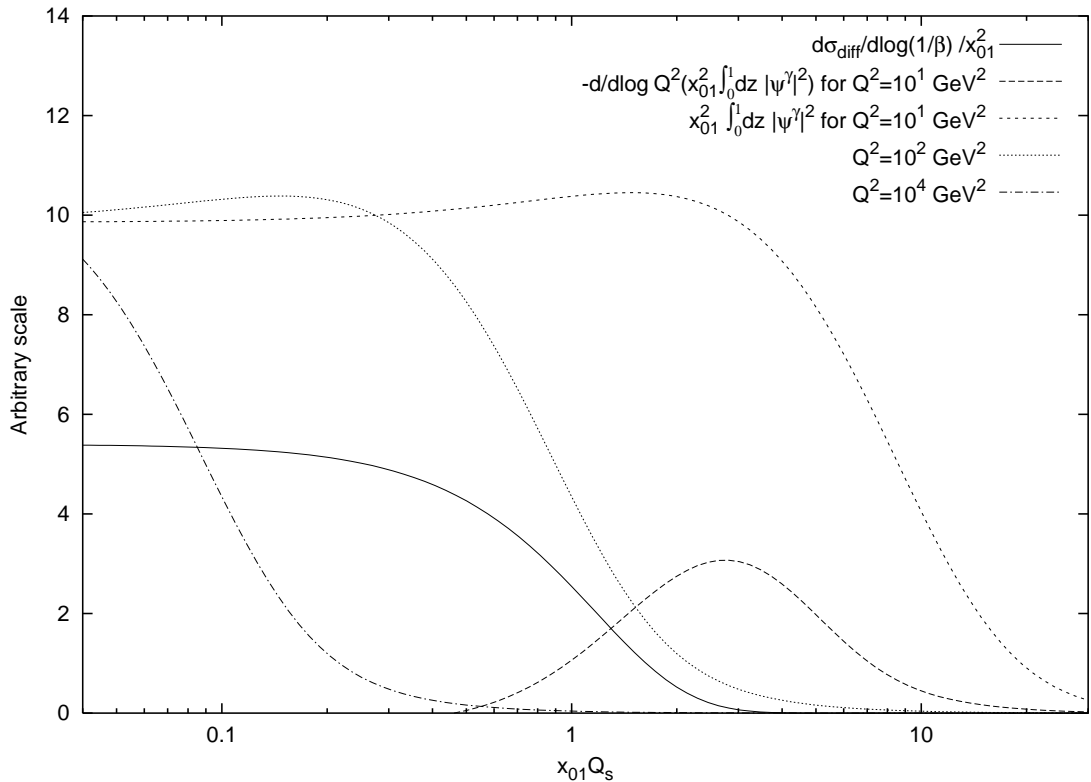,width=15cm}
\caption{\label{fig:plot}The dipole cross section and the photon wave
  function for three different values of $Q^2$. The former is divided
  by the squared size of the dipole, and the latter is multiplied by the
  same, so that the convolution of these two quantities is equal to
  the convolution of the dipole cross section and photon wave function.
The saturation scale $Q_s^2$ is set to $10\ \text{GeV}^2$.
For $Q^2=10\ \text{GeV}^2$, the derivative of the wave function with
  respect to $\log Q^2$ is also shown.}
\end{figure}

As for the case $Q^2 \ll Q_s^2$, one now splits the integral over
$x_{01}$ in
$[0,1/Q_s^2]$, $[1/Q_s^2,4/Q^2]$ and $[4/Q^2,\infty]$.
This time, the first domain gives the main contribution which reads
\begin{equation}
\frac{d\sigmadiff^{\gamma}}{dM_X}=
\frac{\alpha_{em}\alpha_s N_c}{\pi^2 M_X}
\sum e_f^2\frac{1}{Q_s^2}
\int d^2b\,
\int\frac{d^2x}{x^4}(1-S^2(x,b,x_\pom))^2\ ,
\label{eq:approxsmallqcs}
\end{equation}
and which is independent of $Q^2$ and $Q_s^2$ (recall that the
integral over the vector $x$ is proportional to $Q_s^2$).
This can be illustrated graphically.
As one goes to smaller $Q^2$, 
$x_{01}^2\int dz_1|\psi^\gamma(x_{01},z_1~;Q)|^2$ 
stays constant for lower values of
$x_{01}^2$, and develops a plateau \cite{Nikolaev:1994bg}, 
as seen on Eq.~(\ref{eq:approxwf1}) and on 
Fig.~\ref{fig:plot}. 
This plateau contains configurations for which the longitudinal
momentum of the photon is equally shared between the quark and the
antiquark. On the other hand,
$d\sigmadiff/d\log(1/\beta)/x_{01}^2$
is roughly a constant for $x_{01}^2Q_s^2\ll 1$ and drops rapidly to zero for
$x_{01}^2Q_s^2\gg 1$. Hence once $Q^2$ is small enough to allow
the plateau of the photon wave
function to extend up to dipole sizes of the order of
$|x_{01}|\sim 1/Q_s$, the cross section does not
increase anymore when $Q^2$ decreases, and tends to a constant.
The precise value of the constant is model-dependent.

These features are specific to the transverse photon and to the
purely diffractive cross section.

\section{Phenomenology}
\label{sec:3}

We are going to apply the formalism developed above to deep-inelastic
scattering at HERA. Our starting point is Eq.~(\ref{eq:sigmadiffgamma}).

\subsection{From a nucleus to a proton target}
\label{sec:3A}

To go from scattering on a nucleus to scattering on a proton, one
has to take some model assumptions for the non-perturbative physics
inherent to the target and that are not under theoretical control.

First, we assume that the incident dipole is always much smaller than the
target proton, which is reasonable for $Q^2$
not too small compared to $\LambdaQCD^2$.
We also consider a cylindrical target, {\it i.e.} we neglect all 
dependence upon impact
parameter. As far as we do not consider observables which would depend
on the momentum transfer to the proton, this assumption only hampers
predictivity for the global normalization, and brings a
significant technical simplification.

Although we are now dealing with a
proton, we still keep the hypothesis of independence of dipole
interactions in the derivation of Eq.~(\ref{eq:sigmadiffgamma}), which
was {\it a priori} justified for a large nucleus only. 
This could be a problem for $Q^2<Q_s^2$.
But it has recently been shown
(see \cite{Munier:2002gf,Iancu:2002aq} and \cite{Iancu:2001md} for the
theoretical justification) that the proton can be seen
effectively as formed of color neutral domains of typical size
$1/Q_s$. Thus deep in the saturation regime $Q^2\ll Q_s^2$, the proton
is not different from the nucleus, in the sense that the dipoles in
the projectile interact at most once with each of these color
independent domains in the proton.

\subsection{The saturation model}
\label{sec:GBW}

In practice, we choose to stick to the Golec-Biernat and W{\"u}sthoff (GBW) model
\cite{Golec-Biernat:1998js} for the dipole-proton ${\mathbb S}$-matrix, which
reads
\begin{equation}
S(r,b,x_\pom)=\Theta(R_p-b) e^{-Q_s^2(x_\pom) r^2/4}+\Theta(b-R_p)\ ,
\label{eq:GBW}
\end{equation}
where $Q_s^2(x_\pom)=(x/x_0)^{-\lambda}$ in units of $1\ \mbox{GeV}^2$
and $R_p$ is the radius of the proton, related to a normalization
parameter in GBW model
by $R_p^2=\sigma_0/(2\pi)$. $x_0$, $\lambda$ and $\sigma_0$ were
fitted to the data for the total structure function $F_2$ 
in Ref.~\cite{Golec-Biernat:1998js}, we just take
over the found parameters $\sigma_0=23\ \mbox{mb}$, $\lambda=0.288$ and
$x_0=3.04\times 10^{-4}$.
The integration over the impact parameter 
that appears in Eq.~(\ref{eq:sigmadiffgamma})
can be performed, and yields
a factor $\sigma_0/2$.
In addition, in the GBW model, three light quarks were considered, and
a mass of $140\ \mbox{MeV}$ was assigned to them, to ensure a sensible
extrapolation to photoproduction. We take over this feature to our model.

\subsection{A comparison to the data}

We find useful for numerical evaluation to rewrite the 
onium diffractive dissociation cross section as
\begin{multline}
\frac{d\sigmadiff}{d^2b d\log(1/\beta)}=\frac{2\alpha_s C_F}{\pi^2}
\int_0^1 d\lambda_1
\int_0^1 d\lambda_2
\frac{1}{1\!-\!\lambda_1\!+\!2\lambda_1\lambda_2}
\frac{1}{\sqrt{\lambda_2(1\!-\!\lambda_2)
(1\!+\!\lambda_1\lambda_2)(1\!-\!\lambda_1\!+\!\lambda_1\lambda_2)}}\\
\times\left\{\frac{1}{\lambda_1}
\Sigma^2\left(x_{01};\lambda_1;1\!-\!\lambda_1\!+\!2\lambda_1\lambda_2\right)
+\lambda_1\Sigma^2\left(x_{01};
1/\lambda_1;1/\lambda_1\!-\!1\!+\!2\lambda_2\right)\right\}\ ,
\end{multline}
where $\Sigma(x_{01};a;b)=S(a x_{01})S(b x_{01})-S(x_{01})$.
Through an appropriate change of variable, we have mapped the
complex plane into the finite domain $[0,1]\times[0,1]$.
Furthermore, all the factors of the integrant are not more
singular than $1/\sqrt{\lambda}$ for $\lambda\rightarrow 0$,
which is integrable. So this formula does not require numerical
cancellations between large terms, which would result in large
errors. Note also that the obtained formula is quite simple, and this
feature might be related to the conformal invariance of the dipole
splitting kernel.

Putting everything together, the formula that 
we have to evaluate numerically is
\begin{multline}
\frac{d\sigmadiff^\gamma}{dM_X}=
\frac{4\alpha_s N_c}{\pi}\frac{\sigma_0}{M_X} 
\int_0^\infty dx_{01}\,x_{01}
\int_0^1
dz(|\psi_T^\gamma(x_{01},z;Q)|^2+|\psi_L^\gamma(x_{01},z;Q)|^2)\\
\times\int_{[0,1]\times[0,1]} d\lambda_1 d\lambda_2
\frac{1}{1\!-\!\lambda_1\!+\!2\lambda_1\lambda_2}
\frac{1}{\sqrt{\lambda_2(1\!-\!\lambda_2)
(1\!+\!\lambda_1\lambda_2)(1\!-\!\lambda_1\!+\!\lambda_1\lambda_2)}}\\
\times\left\{\frac{1}{\lambda_1}
\Sigma^2\left(x_{01};\lambda_1;1\!-\!\lambda_1\!+\!2\lambda_1\lambda_2\right)
+\lambda_1\Sigma^2\left(x_{01};
1/\lambda_1;1/\lambda_1\!-\!1\!+\!2\lambda_2\right)\right\}\ .
\label{finalforminel}
\end{multline}
where $\Sigma$ is constructed from the ${\mathbb S}$-matrix
element~(\ref{eq:GBW}), and the photon wave functions are given
by Eq.~(\ref{eq:wf}), with flavors and quark masses chosen as
explained in Sec.~\ref{sec:GBW}.

The data do not distinguish between elastic and diffractive
contributions, so we have to add a component in which the final state
is a $q\bar q$ pair. 
This component is not of direct relevance for us as it 
gives a non-negligible contribution for large $\beta\!\sim\! 1$
only. As it has been extensively studied in the literature 
within different models
\cite{Bialas:1996tn,Bialas:1998sb,Bartels:1998ea,
Golec-Biernat:1998js}, we just reproduce the parametrization obtained in 
Ref.~\cite{Golec-Biernat:1998js} to estimate the importance of 
this contribution.

A priori, there is no free parameter. However, by taking $\alpha_s$ at
its value at a phenomenologically realistic 
scale $Q^2\sim Q_s^2\sim 1\ \mbox{GeV}^2$, we overshoot the data by a
factor $1.5-2$. We must set $\alpha_s$ to a lower value
$\alpha_s=0.15$. 

However, the fact that we do not predict correctly the global
normalization does not come as a surprise. Indeed, first, the
(non-perturbative) assumption of a cylindrical target is certainly not
realistic. Second, we took the size of the cylinder over from the
normalization of the cross section fitted to total cross
sections. Because diffractive dissociation is resonant with
the saturation scale, the impact parameter region which
contributes to the cross section should be 
effectively smaller\footnote{We thank Edmond
  Iancu for having drawn our attention to this point.}, which would go
in the right way as far as the normalization is concerned.

\begin{figure}
\epsfig{file=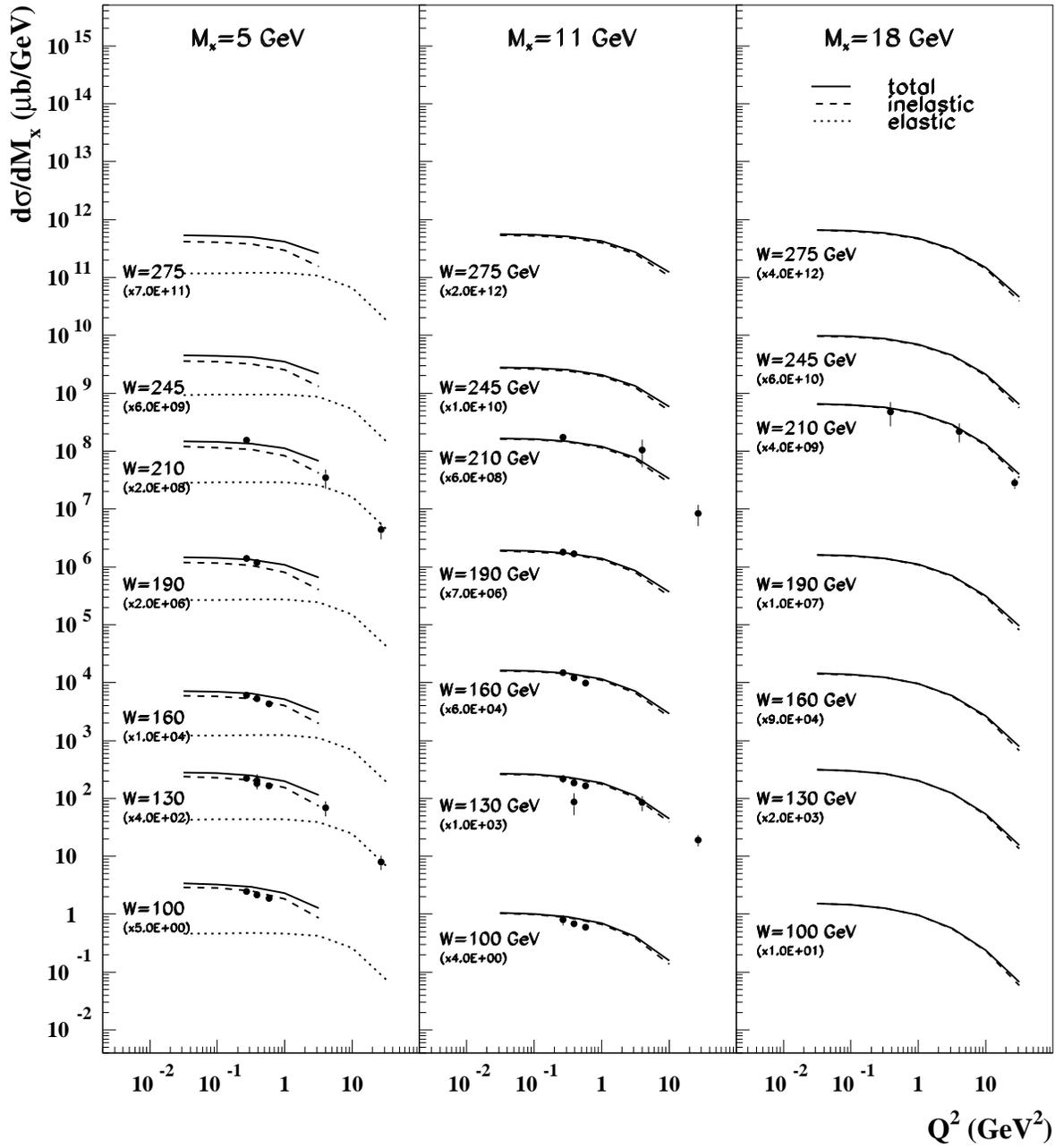,width=18.5cm}
\caption{\label{fig:fig1}Comparison of the model with the data.
The dashed line is our computation of the inelastic
component~(\ref{finalforminel}). 
The dotted line is the elastic component, which is taken from
Ref.~\cite{Golec-Biernat:1998js}. The full line is the sum of these two
components, which for high diffracted masses, reduces to the inelastic
component. The data points are taken from \cite{Chekanov:2002bz}.}
\end{figure}

Our predictions are presented on Fig.~\ref{fig:fig1}. As our model is
established in the limit of small values of $\beta$, 
we choose to restrict our calculation to $\beta<0.2$. This explains why our curves do
not extend to very large $Q^2$, especially at $M_X\!=\!5\ \mbox{GeV}$.
We see that we get a good agreement with the data.

\subsection{Future prospects on phenomenology}

On Fig.~\ref{fig:fig1} we see that the cross section tends to a
constant at low $Q^2$,
as anticipated in Sec.\ref{sec:lim}. This was shown to be due to the
fact that $x_{01}^2\int_0^1 dz\,|\psi_T^\gamma|^2$ also tends to a constant
at low $Q^2$, see Fig.~\ref{fig:plot},
which is unavoidable when the initial state is a
transversely polarized photon. 
The consequence is that the high sensitivity of these diffractive
observables to the exact form of the ${\mathbb S}$-matrix in the saturation
region is somewhat spoiled by this smearing.

However, this property is not true for a
longitudinally polarized photon. Alternatively the plateau at small
$x_{01}^2 Q_s^2$ can also be
eliminated by considering the derivative of the cross section with
respect to $\log Q^2$, namely $d\sigmadiff^\gamma/dM_X
d\log Q^2$ \cite{Nikolaev:1994bg}, see Fig.~\ref{fig:plot}.
A measurement of this quantity would help the
understanding of the ${\mathbb S}$-matrix in the saturation regime, by
enabling a direct scan of this interesting region. A 
more detailed study will be provided elsewhere.


\section{Conclusion}
\label{sec:4}

Diffraction is a good place to study saturation effects in
deep-inelastic scattering.
It had already been shown that quasi-elastic diffractive scattering, like
vector meson production, enables to measure how far one is from the
unitarity limit and
gives a handle on the impact-parameter
dependence of the saturation scale \cite{Munier:2001nr,Rogers:2003vi}.
In this paper, we have provided a theoretical tool 
to study ``true'' (inelastic) diffraction in the saturation regime.
We have formulated the diffractive cross section for a general model
for the dipole-proton ${\mathbb S}$-matrix, and we have rederived an already
known formula in an elegant way using the Good and Walker picture.
When we take for $S$ the Golec-Biernat and W{\"u}sthoff model, we obtain
parameter-free predictions (up to the global normalization) for the
observable $d\sigma/dM_X$ currently measured with good precision at HERA.
An important highlight of our analysis is that diffractive 
observables can help to discriminate in a unique way between the 
predictions of 
different models in the saturation region. This point deserves more
studies, that we leave for the future.

On the theoretical side, the formalism used here based on the dipole
model could be easily
generalized to an arbitrary number of gluons in the final state. This
program has already been explored in
Refs.~\cite{Mueller:1994jq,Bialas:1996bs,Munier:1998nj}
but no unitarity corrections were taken into account.
Our method allows to take into account these corrections.
Whether a simple evolution equation in $\beta$ would be found is not clear.
Anyway, the available energies at present colliders do not require
yet such higher Fock states, nor the complete resummation of all
leading logs, {\it i.e.} of the powers of 
$\alpha_s\log(1/\beta)$.

On the phenomenological side,
it would be worth to take into account more accurately the dependence
upon impact parameter, which is of great importance in the whole
discussion of saturation \cite{Munier:2001nr,Shoshi:2002in}. Therefore, one could
replace the GBW ansatz for the dipole ${\mathbb S}$-matrix
by Kowalski-Teaney \cite{Kowalski:2003hm} recent model, which takes
into account the transverse profile of the proton. 
This replacement might be necessary in order to describe accurately
enough the high precision preliminary data, and
we would have a prediction also for the
normalization of the cross section instead of tuning 
it as we have done here. Furthermore, more exclusive
observables like $d\sigma/dM_Xdt$, 
which will be measured in the future could be predicted.


\appendix
\section{High mass diffraction from the collinear limit}

In this appendix, we compare our approach (see
Eq.~(\ref{eq:approxlargeqcs}))
to the impact factor used by
Golec-Biernat and W{\"u}sthoff for
$q\bar q g$ final states. The latter is formally valid only for
$Q^2\gg Q_s^2$, {\it i.e.} in the collinear limit. 
The two calculations should however lead to the same result when both
$Q^2\gg Q_s^2$ and $\beta\ll 1$, which is the so-called double
logarithmic limit.

According to Ref.~\cite{Bartels:1998ea,Golec-Biernat:1998js},
the diffractive structure function reads
\begin{multline}
\frac{d\sigmadiff^\gamma}{dM_X}
=\frac{81\alpha_{em}\beta M_X}{64\pi^3 B Q^2(Q^2+M_X^2)}
\sum_f e_f^2\frac{\alpha_s}{2\pi}
\int_\beta^1\frac{dz}{z}\left(\left(1-\frac{\beta}{z}\right)^2
+\left(\frac{\beta}{z}\right)^2\right)\frac{z}{(1-z)^3}\\
\times\int \frac{d^2 k}{(2\pi)^2}k^4
\log\left(\frac{(1-z)Q^2}{k^2}\right)\Theta((1-z)Q^2-k^2)
\int d^2r d^2r^\prime
e^{ik(r-r^\prime)}
\sigma_{gg}(r,x_\pom)\sigma_{gg}(r^\prime,x_\pom)\\
\times
\sum_{m,n}(\delta^{mn}-2\frac{r^m r^n}{r^2})
(\delta^{mn}-2\frac{r^{\prime m} r^{\prime n}}{r^{\prime 2}})
K_2\left(\sqrt{\frac{z}{1-z}}k r\right)
K_2\left(\sqrt{\frac{z}{1-z}}k r^\prime\right)
\end{multline}
where $\sigma_{gg}$ is the adjoint dipole-proton cross section, and
$B$ is the exponential $t$-slope of the differential cross section:
\begin{equation}
\frac{d\sigmadiff^\gamma}{dt}=\left.\frac{d\sigmadiff^\gamma}{dt}\right|_{t=0}
e ^{-Bt}\ .
\end{equation}

One can extract the leading
contribution to this cross section in the 
$\beta\rightarrow 0$ limit. 
It comes from the integration region $z\sim\beta$. 
Using $K_2(x)\underset{x\rightarrow 0}{\sim} 2/x^2$ and performing the
sum over helicities $m,n$, one gets
\begin{multline}
\frac{d\sigmadiff^\gamma}{dM_X}
=\frac{81\alpha_{em}\beta}{8 \pi^3 B Q^2 M_X}
\sum_f e_f^2\frac{\alpha_s}{2\pi}
\int_\beta^1\frac{dz}{z^2}\left(\left(1-\frac{\beta}{z}\right)^2
+\left(\frac{\beta}{z}\right)^2\right)
\int \frac{d^2 k}{(2\pi)^2}
\log\left(\frac{Q^2}{k^2}\right)\Theta(Q^2-k^2)\\
\times\int \frac{d^2r}{r^2}\frac{d^2r^\prime}{r^{\prime 2}} e^{ik(r-r^\prime)}
\sigma_{gg}(r,x_\pom)\sigma_{gg}(r^\prime,x_\pom)
\left(
2\frac{(r\cdot r^\prime)^2}{r^2 r^{\prime 2}}-1
\right)
\end{multline}
The integral over $z$ is now factorized and
gives just a factor $2/3\beta$ in the limit
$\beta\rightarrow 0$.
The integral over $k$ can be done analytically
\begin{equation}
\int \frac{d^2k}{(2\pi)^2}\log\left(\frac{Q^2}{k^2}\right)
e^{ik(r-r^\prime)}\Theta(Q^2-k^2)
=\frac{1}{\pi}\frac{1-J_0(Q|r-r^\prime|)}{|r-r^\prime|^2}\ .
\end{equation}
One performs the change of variable 
$R=Q_s r$, $X=Q_s(r-r^\prime)$.
The following expression results:
\begin{equation}
\frac{d\sigmadiff^\gamma}{dM_X}=\frac{27\alpha_{em}}{4 \pi^4 B Q^2 M_X}
\sum_f e_f^2\frac{\alpha_s}{2\pi}
\int d|X|
\frac{1-J_0((Q/Q_s)|X|)}{|X|} f(|X|)\ ,
\label{eq:gbwdiffint}
\end{equation}
where we have defined
\begin{equation}
f(|X|)=Q_s^2\int_0^{2\pi} d\phi\int\frac{d^2R}{R^2(R\!-\!X)^{2}}
\hat\sigma_{gg}(R,x_\pom)\hat\sigma_{gg}(R\!-\!X,x_\pom)
\left(
2\frac{(R\cdot (R\!-\!X))^2}{R^2 (R\!-\!X)^{2}}\!-\!1
\right)\ ,
\end{equation}
and $\phi$ is the angle of vector $X$, $\hat \sigma$ is the dipole
cross section after change of variable. 
$f(|X|)$ is a smooth regular function of $|X|$ having finite value at
$X=0$ and decreasing like $1/|X|^2$. Let us
choose a number $\lambda> 1$ such that $|J_0(\lambda)|\ll 1$. The
integral over $|X|$ can be decomposed as follows:
\begin{equation}
f(0)\left[\int_0^{\lambda Q_s/Q} d|X|
\frac{1-J_0((Q/Q_s)|X|)}{|X|} \frac{f(|X|)}{f(0)}
+\int_{\lambda Q_s/Q}^{1}
\frac{d|X|}{|X|} \frac{f(|X|)}{f(0)}+
\int_1^{\infty} \frac{d|X|}{|X|} \frac{f(|X|)}{f(0)}\right]\ .
\end{equation}
As the integrand is bounded by a number proportional to $Q/Q_s$, 
the first term in the square brackets tends to a constant for $Q\gg Q_s$.
The third term is also a number since $f(|X|)$ decreases sufficiently
quickly to ensure convergence of the integral.
The leading contribution to the second term 
is $\log(Q/Q_s)$.
The coefficient of this logarithm is the value of $f$ at zero, namely
\begin{equation}
f(0)=2\pi\int {d^2r}\frac{\sigma^2_{gg}(r,x_\pom)}{r^4}\ .
\end{equation}
Putting everything together back into Eq.~(\ref{eq:gbwdiffint}), one
finally gets
\begin{equation}
\frac{d\sigmadiff^\gamma}{dM_X}=\frac{27\alpha_{em}}{4\pi^3 B}
\frac{1}{Q^2 M_X}\sum_f e_f^2
\frac{\alpha_s}{2\pi}\log\frac{Q^2}{Q_s^2}
\int {d^2r}\frac{\sigma^2_{gg}(r,x_\pom)}{r^4}\ .
\label{a8}
\end{equation}
In order to compare with our approach, one has to recall (see Sec.\ref{sec:3A}) that
in the large-$N_c$ limit,
$\sigma_{gg}$ is given by
\begin{equation}
\sigma_{gg}(r,x_\pom)=2\int d^2b\,
(1-S^2(r,b,x_\pom))=\sigma_0(1-S^2(r,x_\pom))\ ,
\label{a9}
\end{equation}
where $S$ is the $3-\bar 3$ dipole ${\mathbb S}$-matrix element.
Replacing Eq.~(\ref{a9}) into Eq.~(\ref{a8}), the ratio
between~(\ref{a8}) 
and~(\ref{eq:approxlargeqcs}) reduces to a constant:
\begin{equation}
\frac{3}{4\pi}\frac{\sigma_0}{B}\ .
\end{equation}
The fact that the results match only up to a factor is to be traced to
the different treatments of the integration over impact parameter 
in the two cases.


\begin{acknowledgments}
We thank Yuri Kovchegov for useful correspondence, Edmond Iancu and
Robi Peschanski
for valuable suggestions, Al Mueller for useful discussions and
Bernard Pire for his reading of the manuscript.
This work was started while S.M. was an ESOP Network fellow at the
University of Heidelberg, funded by the
European Commission IHP program under contract HPRN-CT-2000-00130.
He also thanks the Service de physique th{\'e}orique, CEA/Saclay
for hospitality when this paper was being completed.
A.~Sh. is supported by the Deutsche Forschungsgemeinschaft under
contract Sh~92/1-1.
\end{acknowledgments}


\end{document}